\documentclass[11pt,a4,twoside]{article}

\usepackage{./styling/style-main}
\graphicspath{{plots/}}
\usepackage{framed}
\usepackage{amsmath}
\usepackage{slashed}

\makeatletter
\renewcommand\paragraph{\@startsection{paragraph}{4}{\z@}%
	{3.25ex \@plus1ex \@minus.2ex}%
	{-1em}%
	{\normalfont\normalsize\bfseries}}
\makeatother

\usepackage{tikz-feynman}
\usepackage{rotating}
\usepackage{transparent}

\newcommand{\abs}[1]{\vert #1 \vert}

	\title{Propagation of gauge fields in hot and dense plasmas at higher orders}

\author{Andreas~Ekstedt\thanks{andreas.ekstedt@desy.de}\textsuperscript{~~,\,a,\,b}}
\affil{a:~II. Institute of Theoretical Physics, Universität Hamburg, D-22761, Hamburg, Germany}
\affil{b:~Department of Physics and Astronomy, Uppsala University, P.O. Box 256, SE-751 05 Uppsala, Sweden}

\date{\today}
\begin{document}

{\let\newpage\relax\maketitle}
	\thispagestyle{plain}
	\begin{abstract}
		Thermal field theory is indispensable for describing hot and dense systems. Yet perturbative calculations are often stymied by a host of energy scales, and tend to converge slowly. This means that precise results require the apt use of effective field theories. In this paper we refine the effective description of slowly varying gauge field known as hard thermal loops. We match this effective theory to the full theory to two-loops. Our results apply for any renormalizable model and chemical potentials for fermions. We also discuss how to consistently define asymptotic masses at higher orders; and how to treat spectral functions close to the lightcone. In particular, we demonstrate that the gluon mass is well-defined to next-to-leading order.
	\end{abstract}

\section{Introduction}
Thermal field theory is eclipsed in its import only by its complexity. Even so, to describe dense objects like neutron stars~\cite{Vuorinen:2003fs,Gorda:2021kme}, or the production of energetic particles~\cite{Arnold:2001ba,Arnold:1997gh}, we rarely need the full thing. And the labour lightens once we separate the physics at long scales from that at short ones. For at long distances $L\gg T^{-1}$, fields evolve classically; whereas at short scales $L\sim T^{-1}$,  they act like particles~\cite{Arnold:1997gh,Blaizot:1993zk,Arnold:1998cy}. Which in turn means that their response, to an external field, is akin to that of a free electron. So an electric field, for example, can only propagate for so far before it is screened by swift electrons. Physically these electrons start in equilibrium, are perturbed by a field, generate a current, and thereby affect the dynamics of the original field. Therefore the $\mathit{soft}$ field follows effective equations of motion\te known as hard thermal loops~\cite{Braaten:1989mz,Braaten:1990az,Pisarski:1989cs,Braaten:1991gm,Blaizot:1993zk,Frenkel:1989br}.	

To calculate these hard thermal loops we have to integrate out $\mathit{hard}$ modes that have a characteristic energy of order $T$.
 
This can be done by matching correlators in the effective and in the original theory; which, while lending itself to general gauges, becomes arduous at higher orders. So in this paper we sacrifice generality for finesse. In particular, our approach exploits that the influence of soft fields, on hard modes, follows from transport equations~\cite{Blaizot:1993be,Blaizot:1999xk,Blaizot:1993zk}. In effect we calculate the current due to these hard modes\te whose propagation takes place in the soft field's background~\cite{Abbott:1980hw,Abbott:1981ke}. Since such currents can be found with ease, the calculations are simpler, and we are able to incorporate two-loop corrections for any renormalizable model and chemical potential. The downside is that our method is only straightforward in Feynman gauge.
 
That said, it is of course essential to check gauge invariance. Not to mention comparing with other methods. And as luck would have it, an independent calculation has been done for quantum chromodynamics at finite chemical potential~\cite{Gorda:2022xx}. This calculation uses a general gauge and complements the approach taken in this paper\footnote{Results also exist for quantum electrodynamics~\cite{Gorda:2022fci,Carignano:2019ofj,Carignano:2017ovz} at two loops. }. 

Yet even with hard thermal loops in hand, it can be difficult to extract consistent results close to the lightcone\te when the soft photon or gluon is almost on-shell. This is because a hard electron can absorb such a photon and travel a great distance before it is reemitted. However, we know that physical degrees of freedom pick up thermal masses as they propagate~\cite{Weldon:1982aq,Weldon:1982bn,Klimov:1982bv}, which means that hard electrons can only travel for so far, and further suggests that we should resum our propagators~\cite{LEBEDEV1990229,Kraemmer:1994az}. We confirm this by showing that ill-behaved terms at two-loops are the very same that get resummed. After using these resummations we can then consistently extract asymptotic masses for gluons and photons.

\section{Hard thermal loops at finite temperature and density}
Let us now understand how hard thermal loops arise at higher orders. The details are given in~\cite{Ekstedt:2023anj}, and we here focus on the physics. We use a mostly-plus metric and denote four-vectors by  $K^2=-(k^0)^2+\vec{k}^2$ and spatial vectors by $k^2\equiv \vec{k}^2$. In addition, all calculations assume Feynman gauge.

Consider quantum electrodynamics. The idea is to integrate out hard $E\sim T$ positrons and electrons, and to describe how soft fields propagate. To do this we use transport equations~\cite{Blaizot:1993zk,Blaizot:1993be,Blaizot:2001nr}:
\begin{align}
	\dot{N}_F^{\pm}+\vec{v}\cdot \vec{\nabla}N_F^{\pm}\pm e \left(\vec{E}+\vec{v}\times \vec{B}\right)\cdot \vec{\nabla}^p N_F^{\pm}=0, \quad \vec{v}=\vec{p}/p^0.
\end{align}
By assumption the electric field varies slowly, and we can linearize around equilibrium:
\begin{align}\label{eq:Vlasov}
	&N_F^{\pm}=\frac{1}{2}-n^{\pm}_F-\delta n_F^{\pm}, \quad v\cdot \partial \delta n_F^{\pm}=\mp e \vec{v}\cdot \vec{E} \partial_{p^0} n_F^{\pm},
\end{align}
where $n_F^{\pm}(p,\mu_F)=\left(e^{p/T\pm \mu_F/T}+1\right)^{-1}$, and $\mu_F$ is the electron's chemical potential. 

As these charges propagate they will in turn generate a current:
\begin{align}
	\partial_\mu F^{\nu \mu}=2 e \int \frac{d^3p}{(2\pi)^3} v^{\nu} \left[N^{+}_F-N^{-}_F\right],\quad v^\nu=(1,\vec{v}).
\end{align}
Without a chemical potential the plasma is neutral, and so no current arises in equilibrium. Yet since electric fields affect positive and negative charges differently, a current does arise from deviations about equilibrium:
\begin{align}\label{eq:LOHTL}
	\partial_\mu F^{\nu \mu}=-2 e \int \frac{d^3p}{(2\pi)^3} v^{\nu} \left[\delta n^{+}_F-\delta n^{-}_F\right]=-\frac{e^2 T^2}{3}\int \frac{d\Omega_v}{4\pi} \frac{v^\nu \vec{v}\cdot \vec{E}}{v\cdot K}\equiv \Pi^{\mu \nu}A_\mu.
\end{align}
Physically what happens is that electrons with velocity $\vec{v}$ are perturbed by the electric field, and travel on average $t\sim \left[(p^0+k^0)-\abs{\vec{p}+\vec{k}}\right]^{-1}\sim (v\cdot K)^{-1}$ before their influence is felt. In that time the electrons have changed their momentum by $\delta \vec{p}\sim t e \vec{E}$, and thus provide a net current $\delta \vec{j}\sim e n_F (\delta \vec{v})\sim  t \left(T^2 e^2  \vec{E}\right)$.

In any case, from equation \ref{eq:LOHTL} we identify the hard-thermal-loop  self energy:
\begin{align}\label{eq:HTLLO}
\Pi_\text{LO}^{\mu \nu}=-\frac{e^2 T^2}{3} \Pi_1^{\mu \nu}.
\end{align}
where 
\begin{align}
	&\Pi_1^{\mu \nu}=\int\frac{d\Omega_v}{4\pi}\left(k^0 \frac{v^\mu v^\nu}{v\cdot K}+n^\mu n^\nu\right), \quad n^{\mu}=\left(1,\vec{0}\right).
\end{align}
We can likewise understand how two-loop contributions arise. In particular, as the electrons travel, they interact with the plasma and pick-up thermal masses $m^2_F\approx \frac{1}{4}e^2 T^2$; which reduce their speed by $\frac{m^2_F}{2 (p^0)^2}\sim e^2$. After using this reduced speed in equation \eqref{eq:HTLLO} we find the Lorentz structure
\begin{align}
	 \Pi_2^{\mu \nu}=\int \frac{d\Omega_v}{4\pi}\left\{v^\mu v^\nu \left[\frac{(k^0)^2}{(v\cdot K)^2}-\frac{2 k^0}{v\cdot K}\right]+\left[v^\mu n^\nu+n^\mu v^\nu\right]\frac{ k^0}{v\cdot K}-n^\mu n^\nu\right\}.
\end{align}
Non-abelian fields can also exchange colour with the plasma. This does not change their velocities, but does alter their distributions. As such, this contribution is proportional to $\Pi_1^{\mu \nu}$.

If we now introduce a chemical potential, we find a current even in equilibrium:
\begin{align}
	\partial_\mu F^{\nu \mu}\sim e \mu_F(\mu_F^2/\pi^2+T^2) n^\nu+\ldots.
\end{align}
Other than this, the main change is that integrals of Fermi-Dirac distributions now give $T^2+3 \mu_F^2/\pi^2$ where before they gave $T^2$.

 There is however one new Lorentz structure that appears. To see this, note that in a neutral plasma both the equilibrium current and the average deviation $\sim \int_p \left(\delta n^{+}_F+ \delta n^{-}_F\right)$ vanish. Yet once the vacuum is charged, an electron in equilibrium can generate a current, scatter, and travel for $L\sim (v\cdot K)^{-1}$ before interacting with $\delta n^{\pm}$. This process gives rise to a third and final Lorentz structure:
\begin{align}
\Pi_3^{\mu \nu}=\int \frac{d\Omega_v}{4\pi}\left[\frac{k^0 v^\mu}{v\cdot K}+n^\mu\right]\int \frac{d\Omega_w}{4\pi}\left[\frac{k^0 w^\nu}{w\cdot K}+n^\nu\right].
\end{align}

Note that all three Lorentz structures are conserved: $\Pi_i^{\mu \nu}K_\nu=0$. This means that current conservation also holds to two-loops; at least for the hard contribution in Feynman gauge.

\subsection{Other terms in the effective theory}
When integrating out hard particles we are matching our full theory to an effective one at some scale $\mu\sim T$.  For a non-abelian theory we find the effective operator~\cite{Pisarski:1990zj,Braaten:1991gm,Frenkel:1989br}
\begin{align}
j^\mu \sim C^{ab}\int \frac{d\Omega_v}{4\pi} \left[\frac{v^\mu \vec{v}\cdot \vec{E} }{v\cdot D}\right]^b,
\end{align}
where the covariant derivative is $\left[D_\mu N\right]^a=\partial_\mu N^a+g^{abc}A_\mu^b N^c$. At the next order the operator coefficient, $C^{ab}\sim \delta^{ab} T^2 g^2$, is modified such that $C^{ab}$ is independent of the matching scale.

However, when we integrate out hard modes the gauge couplings also change; which means that $g^2\rightarrow g^2_\text{eff}= g^2+a g^4$, where $a$ ensures that $ g^2_\text{eff}$ is independent of the hard matching scale\footnote{This effective coupling coincides with the one in dimensional reduction~\cite{Kajantie:1995dw} up to a factor of $T$.}. So when we write down operators in the effective theory, all gauge coupling appearing in the covariant derivative better be $g_\text{eff}$. This ensures that the effective theory is gauge and matching-scale invariant.

\subsection{Model parametrization}\label{sec:Generic}
We follow \cite{Martin:2018emo,Machacek:1983tz,Machacek:1983fi,Machacek:1984zw} and parametrize our model with generalized couplings. The details are given in \cite{Ekstedt:2023anj} and references therein. In short, $Y^{iIJ}$ are Yukawa couplings, $\lambda^{ijkl}$ are scalar quartics, $g^a_{ij}$ are scalar-vector couplings, $g^{aI}_J$ are fermion-vector couplings, and $g^{abc}$ are vector trilinear couplings. Finally, we denote fermion chemical potentials by $\mu_{F,J}^{I}\equiv \delta^I_J \mu_F^{(I)}$. We do not consider scalar chemical potentials in this work. See appendix \ref{app:FiniteChemical} for new contributions with $\mu_F\neq 0$.

\subsection{Field redefinitions}
As with all matching calculations, our result can\te and will\te depend on the hard gauge-fixing parameter. Normally this gauge dependence cancels for physical quantities. Alternatively one can remove this spurious gauge dependence directly by using a background-field gauge~\cite{Abbott:1980hw,Abbott:1981ke}. An equivalently procedure  is to shift our soft fields such that he kinetic term is canonical. In our case, transverse and longitudinal fields behave differently, so we have two choices: We can either shift all fields by the same factor. Or we can make both transverse and longitudinal fields canonical, and by doing so introducing a new Lorentz structures ($\sim n^\mu n^\nu$) at two loops. 

We opt for the first option; so we choose to only make longitudinal fields canonical (when $k^0$). As such the\te gauge-invariant\te Debye mass can directly be read of from the longitudinal self-energy in the $k^0\rightarrow 0$ limit.

\subsection{Lorentz structures}\label{eq:TwoLoopRes}
Three Lorentz structures appear at two loops:
\begin{align}\label{eq:LorentzStruct}
	&\Pi_1^{\mu \nu}=\int\frac{d\Omega_v}{4\pi}\left(k^0 \frac{v^\mu v^\nu}{v\cdot K}+n^\mu n^\nu \right), \quad n^{\mu}=\left(1,\vec{0}\right), \nonumber
	\\&  \Pi_2^{\mu \nu}=\int \frac{d\Omega_v}{4\pi}\left\{v^\mu v^\nu \left[\frac{(k^0)^2}{(v\cdot K)^2}-\frac{2 k^0}{v\cdot K}\right]+\left[v^\mu n^\nu+n^\mu v^\nu\right]\frac{ k^0}{v\cdot K}-n^\mu n^\nu\right\},
	\\&  \Pi_3^{\mu \nu}=\int \frac{d\Omega_v}{4\pi}\left[ \frac{k^0 v^\mu}{v\cdot K}+n^\mu\right]\int \frac{d\Omega_w}{4\pi}\left[\frac{k^0 w^\nu}{w\cdot K}+n^\nu\right]. \nonumber
\end{align}
All structures are conserved, and $\Pi_3^{\mu \nu}$ only shows up if a chemical potential is present.

\subsection{Contributions at finite chemical potential}\label{sec:HTLTwoloop}
The addition of chemical potentials only modifies fermion contributions. For example, at leading-order we find
\begin{align}
\delta \Pi^{\mu \nu}_F=-\frac{1}{6}\text{Tr}\left[g_F^a\left( 3\mu_F^2/\pi^2\right) g_F^b\right]  \Pi^{\mu \nu}_1+\mathcal{O}(\epsilon).
\end{align}

The full two-loop result is given in appendix \ref{app:FullResult}, where in addition to the terms given in~\cite{Ekstedt:2023anj}, a finite chemical potential generates the terms shown in appendix \ref{app:FiniteChemical}. The calculation of these terms is identical to those in~\cite{Ekstedt:2023anj}; with the exception of the terms proportional to $\Pi_3^{\mu \nu}$, whose derivation is given in appendix \ref{app:Pi3Terms}.

\subsection{Power corrections from one-loop diagrams}\label{sec:PowerCorrection}
Power corrections describe, amongst other things, how fields in the effective theory are related to $T=0$ ones. The extra contributions that arise with a finite chemical potential is given in~\cite{Gorda:2022fci}, which in our notation look like
\begin{align}
	&g_{\mu \nu }\delta \Pi^{\mu \nu}_{F}=\frac{4}{3}\text{Tr}\left[g_F^a  \overline{\aleph}_Fg_F^b\right]\frac{K^2}{16 \pi^2},\quad  \delta \Pi^{00}_F=-\frac{8}{3}\text{Tr}\left[g_F^a \overline{\aleph}_F g^b\right]\frac{k^2}{16 \pi^2}.
\end{align}
Here $\overline{\aleph}_F$ is given in terms of digamma functions  (see the definition of $\aleph(z)$ in \cite{Vuorinen:2003fs})
\begin{align}\label{eq:NF}
\overline{\aleph}_F=-\left\{\frac{1}{2}\left[\psi\left(1/2-i \frac{\mu_F}{2\pi T}\right)+\psi\left(1/2+i \frac{\mu_F}{2\pi T}\right)\right]+\log(4)+\gamma\right\}.
\end{align}
For small $\mu_F\ll T$,
\begin{align}
	\overline{\aleph}_F=-\frac{7 \mu_F^2}{4 \pi^2 T^2}\zeta(3)+\frac{31 \mu_F^4}{16\pi^4 T^4}+\mathcal{O}\left(\mu_F^6\right),
\end{align} 
and for large $\mu_F \gg T$
\begin{align}
\overline{\aleph}_F= \log \frac{\pi T e^{-\gamma}}{2 \mu_F }+\frac{\pi^2 T^2}{6 \mu_F^2}+\mathcal{O}\left(\mu_F^{-4}\right).
\end{align}
\subsection{Result for quantum chromodynamics}\label{sec:QCD}
Consider now quantum chromodynamics with $N_q$ fundamental fermions\te all with the same chemical potential $\mu_F$. We find the two-loop contribution\footnote{
	This result is in agreement with~\cite{Gorda:2022xx}. Note that $k^0 L[K] \Pi^{\mu \nu}_1$ terms can be recast as power corrections, but we keep them here to illustrate their origin.}
\begin{align}\label{eq:QCDTwoLoop}
	\Pi^{\mu \nu}_\text{two-loop}=&-\alpha_s^2 C_A^2 T^2 \left\{\frac{4}{3}\left[\log \frac{\mu e^\gamma }{4 \pi  T}+k^0 L[K] \right]\Pi_1^{\mu \nu}-\frac{2}{3} \Pi_2^{\mu \nu}\right\} \nonumber
	\\&-\alpha_s^2 N_q \left\{-\frac{2 \mu_F^2}{\pi^2}(2C_F T_F-C_A T_F)\Pi_3^{\mu \nu} +2 C_F T_F \left(\frac{\mu_F^2}{\pi ^2}+T^2\right)\Pi_2^{\mu \nu}\right\}
	\\&- \alpha_s^2 N_q \left\{\frac{4}{3} C	_A T_F \left(\frac{3 \mu_F^2}{\pi ^2}+T^2\right) \left[\log \frac{\mu e^\gamma }{4 \pi  T}+k^0 L[K] \right] \Pi_1^{\mu \nu}-\frac{2}{3} C_A T_F \left(\frac{3 \mu_F^2}{\pi ^2}+T^2\right)\Pi_2^{\mu \nu}	\right\}, \nonumber
\end{align}
where the group invariants are defined in appendix \ref{app:GroupInvariants}, and $\mu\sim T$ is the matching-scale. 

What's more, we also use a convention where our fields are shifted to make the longitudinal propagator canonical when $k^0=0$. This shift gives the terms in appendix~\ref{app:QCDFieldShift}.

Finally, we have to add remaining power corrections. For the transverse self-energy we find (see equation \eqref{eq:NF} for the definition of $\overline{\aleph}_F$ )
\begin{align}\label{eq:QCDPower}
	\Pi^T_\text{pow}=&\frac{\alpha_s N_q T_F K^2}{6 \pi}\left\{\left(2-\frac{k_0^2}{k^2}\right)(1-k^0L[K])+5 k^0 L[k]+4 \overline{\aleph}_F \right\}\nonumber
	\\&+\frac{\alpha_s C_A K^2}{12 \pi} \left\{\left(\frac{k_0^2}{k^2}-2\right) (1-k^0L[K])-11 k^0 L[K]\right\}.
\end{align}

The complete next-to-leading order result is then
\begin{align}
\Pi^T_\text{NLO}=\Pi^T_\text{two-loop}+\Pi^T_\text{field shift}+\Pi^T_\text{pow},
\end{align}
where the relations in appendix \ref{app:TL} can be used to find $\Pi^T_\text{two-loop}$ and $\Pi^T_\text{field shift}$. This result, together with the leading-order one, is independent of the renormalization scale to $\mathcal{O}(\alpha_s^2)$ if we use leading-order beta functions.

\section{Derived quantities}
Let us now use our results to study some actual physics. To save ink we set the chemical potential to zero unless specified otherwise.

\subsection{Asymptotic masses at higher orders}
Photons and gluons have two physical modes at high energies. This still holds at finite temperature, but these $\mathit{transverse}$ modes now gain a thermal mass. Take for example electrodynamics; if we study the transverse propagator, we find a pole at $ (k^0)^2-k^2\sim e^2 T^2$ for $ k\gg eT$~\cite{Pisarski:1989cs,Weldon:1982aq,Weldon:1982bn,Klimov:1982bv}. It is natural to associate this pole with a mass since the residue on said pole behaves as $(2k)^{-1}$. 

However, at the next order one finds~\cite{Gorda:2022fci}
\begin{align}
(k^0)^2=k^2+\underbrace{(e^2 T^2+k^{-2}\log k^2)}_\text{From LO}+\underbrace{e^4 T^2\left(1+\log k^2\right)}_\text{From NLO}+\ldots, \quad  k \gg e T,
\end{align}
which hardly resembles a physical mass. In fact, our peculiar logarithms come from 
\begin{align}
\int \frac{d\Omega_v}{4\pi} \frac{1}{v\cdot K}\sim \log\frac{k^0+k+0^{+}}{k^0-k+0^{+}}\sim \log \frac{k^2}{K^2}.
\end{align}
 At leading order this term multiplies $K^2\sim e^2 T^2$, which means that our logarithms are pushed to the next order\te no such mitigation occurs at two loops. 

Yet prodigious guise alone does not dissuade us.

In particular, we can understand what's going on by recalling that these logarithms arise because hard particles can travel a distance $L\sim \left[(p^0+k^0)-\abs{\vec{p}+\vec{k}}\right]^{-1}\sim (v\cdot K)^{-1}$ before affecting the gauge fields. And while these particles can travel far, they don't propagate freely, and $L$ is cut off by thermal masses. So even when $K^2\approx 0$, the maximum distance is of order $L\sim (e^2 T)^{-1}$.  This means that as we approach the lightcone we have to use 
\begin{align}
v\cdot K= -k^0+\vec{k}\cdot\vec{p} \left(p^2+m_F^2\right)^{-1/2}.
\end{align}
 Away from the lightcone it is proper to expand in powers of $m_F$; and by doing so we but recover our two-loop result. However, as we approach $K^2\sim e^2 T^2$, we have to use the full propagator as advocated in~\cite{Kraemmer:1994az,LEBEDEV1990229}. 

All in all, after we resum the relevant contributions we can use $\Pi^T(K)$ to solve $K^2+\Pi^T(K)=0$:
\begin{align}
(k^0)^2\vert_{k\gg e T}=k^2+m_\infty^2+\mathcal{O}(k^{-2}),
\end{align}
where the asymptotic photon mass is\footnote{We have here shifted our fields such that the $A^0 A^0$ propagator is canonical when $k^0=0$; other conventions just move pieces over to power corrections.}
\begin{align}\label{eq:PhotonNLOMass}
m_\gamma^2=\frac{e^2 T^2}{6}-\frac{e^4T^2}{36\pi^2}\left[ \log \frac{\mu e^\gamma}{4\pi T}-\frac{1}{2}+2 \log 2\right]-\frac{1}{4\pi^2} e^2 m_F^2+\mathcal{O}(e^6),\quad m_F^2=\frac{e^2 T^2}{4}.
\end{align}
Note that the mass in equation \eqref{eq:PhotonNLOMass} is $\mu$ independent to $\mathcal{O}(e^4)$ once we use leading-order beta functions. It can also be checked from the results in~\cite{Gorda:2022xx } that this mass does not depend on the gauge-fixing parameter.

That leaves power corrections. Since these are not, as yet, know to two loops we can not confirm that they are resummed. But physically we expect that they will be. As discussed in appendix~\ref{app:SelfEnergy}, resumming these amounts to replacing factors of $K^2 k^{0}L[K]$ by $-m_\gamma^2$; this gives an extra $-\frac{5e^2}{24 \pi^2}\left(m^2_\gamma\right)_\text{LO}$ in equation \eqref{eq:PhotonNLOMass}. The full result is shown in figure \ref{fig:OmegaT}.

\subsection{Asymptotic gluon mass}
At leading-order the gluon self-energy is
\begin{align}\label{eq:OneLoop}
\Pi^{\mu \nu}_\text{LO}=-\frac{4 \pi}{3} \alpha_s C_A T^2 \Pi_1^{\mu \nu}-\frac{4 \pi}{3} \alpha_s T_F N_q \left( T^2+3\frac{\mu_F^2}{\pi^2}\right) \Pi_1^{\mu \nu},
\end{align}
where the first term comes from vectors, and the second from fermions. 

The two-loop contribution is given in equation \eqref{eq:QCDTwoLoop}, and as discussed, $\Pi_2^{\mu \nu}$ terms arise because hard particles pick up thermal masses. We can check this by taking $p^0\rightarrow p+\frac{m_i^2}{2 p}$ in equation \eqref{eq:OneLoop}. Doing so we recognise well-known thermal masses for gluons and quarks\footnote{When doing this we have to re-express $T^2$ and $\mu_F^2$ terms as integrals over the relevant distribution, and then let $p^0\rightarrow p+\frac{m_i^2}{2 p}$ in said distribution}:
\begin{align}
&\left. m_g^2 \right\vert_\text{LO}=\frac{1}{6}g_s^2 T^2\left(C_A+N_q T_F\right) +\frac{1}{2}g_s^2N_q T_F \frac{\mu_F^2}{\pi^2},
\\& \left. m_q^2\right\vert_\text{LO}=\frac{1}{4}C_F g_s^2\left(T^2+\frac{\mu_F^2}{\pi^2}\right).
\end{align}

To find the asymptotic mass at the next order, we should resum all $\Pi_2^{\mu \nu}$ terms into equation \eqref{eq:OneLoop}; keep $\Pi_1^{\mu \nu}$ and $\Pi_3^{\mu \nu}$ terms; add remaining power corrections; and solve $K^2+\Pi_T(K)=0$ in the limit $k, k^0\gg g_s T$. We find
\begin{align}
m^2_\infty=m^2_g\vert_\text{LO}+m^2_g\vert_\text{NLO}, \quad m^2_g\vert_\text{NLO}&=m^2_\text{resum}+m^2_\text{ren}+m^2_\text{finite}+m^2_\text{power},
\end{align}
where we have split contributions according to their origin. In particular, $m^2_\text{pesum}$ comes from resumming two-loop terms; remaining two-loop terms (and field shifts) are split into  $m^2_\text{ren}$ and $m^2_\text{finite}$. Finally, contributions from power-corrections are given by $m^2_\text{power}$. The expressions are given in appendix \ref{app:QCDMassRes}, and figure \ref{fig:GluonMass} shows that for $N_q=6$ the next-to-leading order mass is roughly $30\%$ smaller than the leading-order one.

It is however important to remember that this result only contains contributions from hard $E\sim T$ modes\te to find the full mass we have to calculate the self-energy within the effective theory~\cite{Ghiglieri:2021bom,Caron-Huot:2008vbk,Kraemmer:1994az}.

\begin{figure}
	\begin{center}
		\hspace*{-1cm}
	\includegraphics[width=0.9\textwidth]{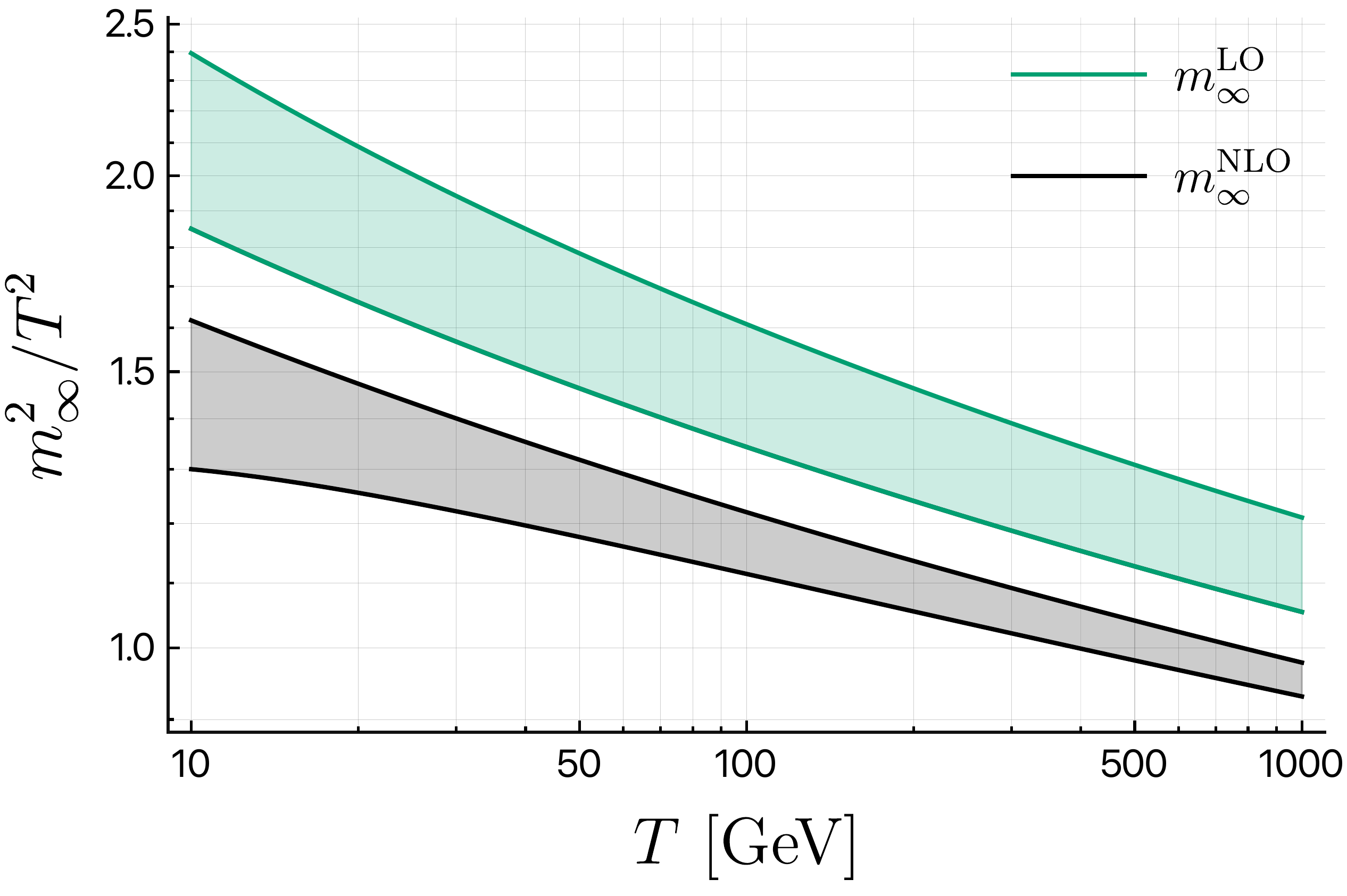}
	\caption{Comparison of the leading-order and the next-to-leading order asymptotic gluon mass. The bands corresponds to varying the renormalization-scale from $\mu=T/2$ to $\mu=2 T$. We use $\alpha_s(M_z)=0.1179$, $\mu_F=0$, and $N_q=6$. All couplings are evolved with one-loop beta functions.}
	\label{fig:GluonMass}
\end{center}
\end{figure}

\subsection{Two-loop correction to the gluon damping rate}
When describing non-perturbative processes, like sphaleron transitions~\cite{Arnold:1999uy,Arnold:1998cy,Arnold:1996dy}, it is important to understand how gauge fields are screened. The damping rate can be defined from the imaginary part of the self-energy~\cite{PhysRevD.47.5589,Pisarski:1989cs}, which essentially coincides with the linearized collision operator~\cite{Arnold:1999uy,Arnold:1998cy}.

 In particular, for transverse gauge bosons the self-energy is of the form (when $k \gg k^0$)
\begin{align}
\Pi^T \approx -i \frac{\pi}{4} \frac{k^0}{k} \delta^2+\mathcal{O}\left(k_0^2\right),
\end{align}
and the gluon damping is of order~\cite{Selikhov:1993ns,Arnold:1999uy,Arnold:1998cy,Bodeker:1999ey}
\begin{align}
\gamma \sim C_A T \alpha_s \frac{m_D^2}{\delta^2} \log \frac{\delta}{\mu}.
\end{align}

Here $m_D^2=\frac{1}{3} g_s^2 T^2(C_A+ N_q T_F)$ is the one-loop Debye mass, which coincides with $\delta^2$ at leading order, while the next-to-leading-order contribution is
\begin{align}\label{eq:GluonDamping}
\left. \frac{\delta^2}{\alpha_s^2 T^2}\right\vert_\text{NLO}=&\frac{2}{9} \left(11 C_A - 4 N_q T_F\right) \left(C_A + N_q T_F\right) \log \frac{\mu e^\gamma}{4\pi T}-\frac{2}{9} \left(C_A + N_q T_F\right) \left(11 C_A + 4 N_q T_F\right)\log 2 \nonumber
\\&+\frac{1}{9} \left\{11 C_A^2+15 C_A N_q T_F+4 N_q T_F \left(N_q T_F-9 C_F\right)\right\}-\frac{2}{9} \left(C_A - 2 N_q T_F\right) \left(C_A + N_q T_F\right).
\end{align}

\begin{figure}
	\begin{center}
		\hspace*{-1cm}
		\includegraphics[width=0.8\textwidth]{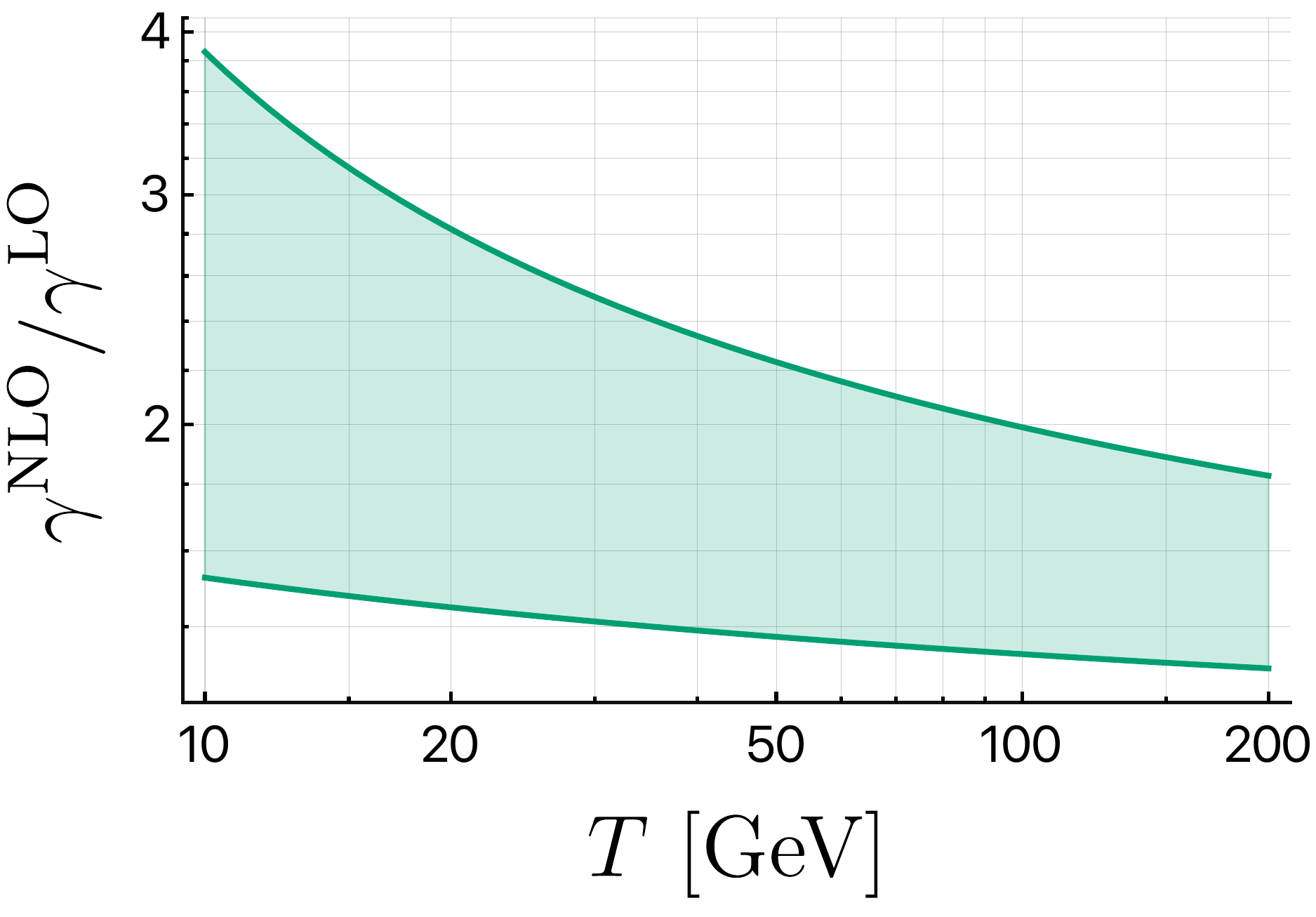}
	\end{center}
	\caption{Rough estimate of the next-to-leading order gluon damping. The bands corresponds to varying the renormalization scale from $\mu=T/2$ to $\mu=2 T$. We use $\alpha_s(M_z)=0.1179$, $\mu_F=0$, and $N_q=6$. Couplings are evolved with one-loop beta functions. In this comparison we use the rough approximation $\frac{\gamma^\text{NLO}}{\gamma^\text{LO}}\approx \frac{m_D^2}{\delta^2}$, where the NLO contribution for $\delta^2$ is given in equation \eqref{eq:GluonDamping}. Note that most scale dependence comes because $m_D^2$ is evaluated at leading order.}
\label{fig:GluonDamping}
\end{figure}

We see in figure \ref{fig:GluonDamping} that next-to-leading order corrections boost $\gamma$ by roughly a factor of 2. Although, this comparison is not completely honest since we have neglected higher-order corrections from the collision operator. Hence figure \ref{fig:GluonDamping} should only be seen as a rough estimate.

\subsection{Spectral functions}
Let us now study the transverse propagator a bit closer:
\begin{align}
\Delta^T(K)=\frac{1}{K^2+\Pi^T(K)}.
\end{align}
If $k^0\geq k$ the propagator has a pole. Furthermore, the self-energy has a branch cut if $k^0< k$; this corresponds to the field being damped and losing energy to hard charges. In total the imaginary part of the propagator is~\cite{PhysRevD.47.5589,Pisarski:1989cs,Laine:2016hma}
\begin{align}
\text{Im}\Delta^T=2\pi \rho^\text{res}(k^0,k)\delta\left(k^0-\omega_T(k)\right)+\rho^\text{disc}(k^0,k)\theta\left(k-k^0\right).
\end{align}
Consider first the spectral density for the pole which is given by $\rho^\text{res}=\left\vert 2k^0-\partial_{k^0} \Pi^T \right\vert^{-1}$. Now, the transverse poles have two important properties that justify us interpreting them as physical excitations. That is, $\omega_T^2\rightarrow k^2+m_\infty^2$ for large $k\gg T$ momenta while the residue scales as $\rho^\text{res}\sim \frac{1}{2k}$ in the same limit.

\begin{figure}
	\begin{center}
		 \hspace*{-1cm}
\includegraphics[width=1.1\textwidth]{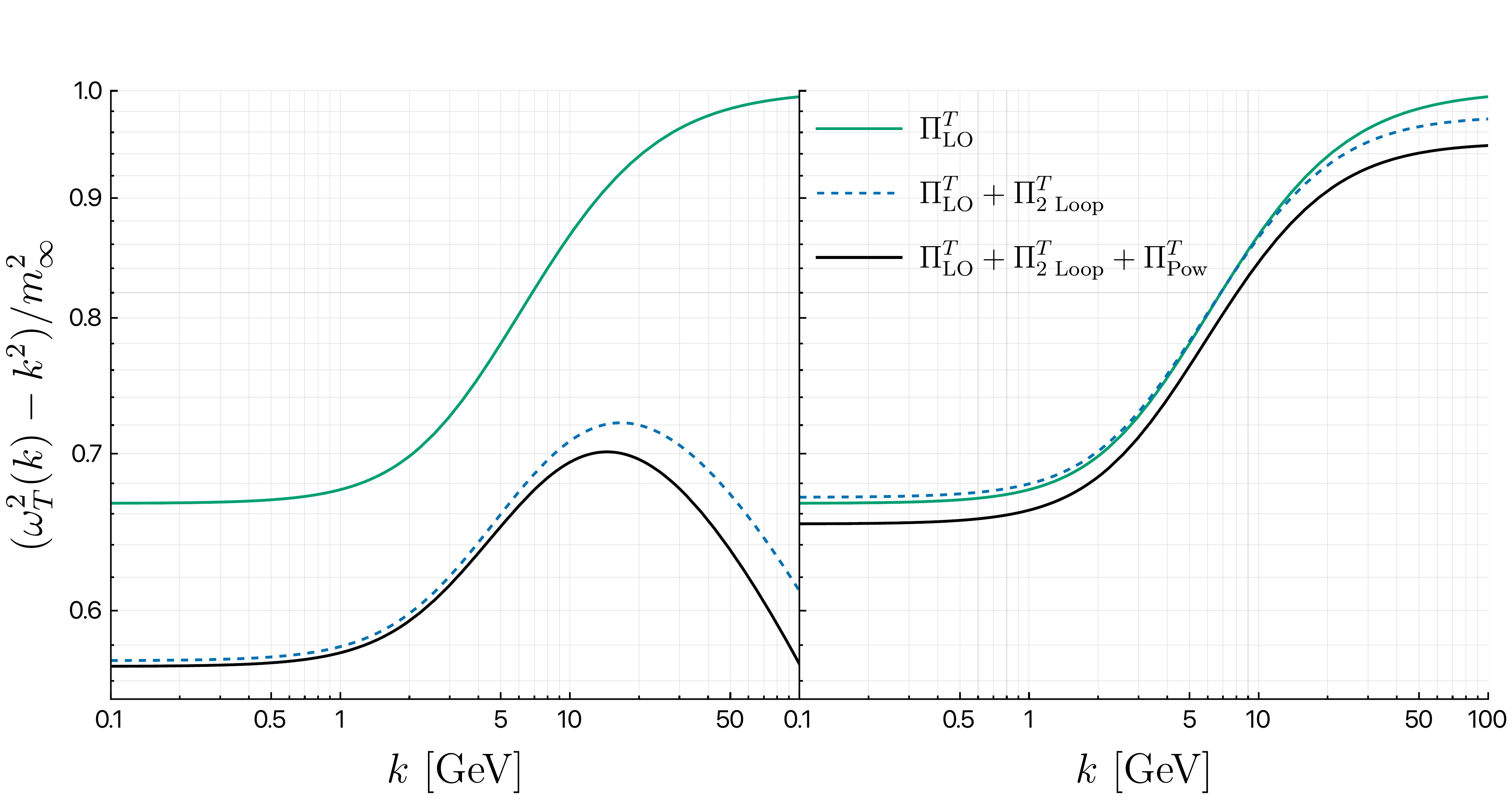}
\end{center}
\caption{Comparison of bare and resummed self-energies at next-to-leading order. In particular the curves show $\left(\omega_T^2(k)-k^2\right)/m_\infty^2$ where we use the leading-order mass in the denominator for all curves. The left plot shows the result without resummations, whereas the right plot shows the same comparisons with resummations. Note that for the right plot $\Pi^T_\text{LO}+\Pi^T_\text{2 Loop}$ denotes that two-loop terms have been resummed. In addition, for the right plot the result has been expanded for $k\geq T$, and the small-$k$ region is not accurate. We use $\alpha=0.1$ and $\mu=T=10~\text{GeV}$ in both plots.}
\label{fig:OmegaT}
\end{figure}

Let us now see how these quantities behave for electrodynamics; we consider one Dirac fermion with zero chemical potential. The result does not qualitatively depend on the details, so as a benchmark we take $\mu=T=10~\text{GeV}$ and $\alpha(T)=\frac{e^2(T)}{4\pi}=0.1$. We see from figure \ref{fig:OmegaT} that higher-order corrections rapidly run rampant as $k$ grows\footnote{We do not solve for the pole in a strict perturbative expansion around the leading-order pole.}. The resummed result, on the other hand, behaves well for large $k$ and approaches $\omega_T^2-k^2\approx m_\infty^2$ as $k\gtrsim T$.  

\begin{figure}
	\begin{center}
		\hspace*{-1cm}
		\includegraphics[width=0.9\textwidth]{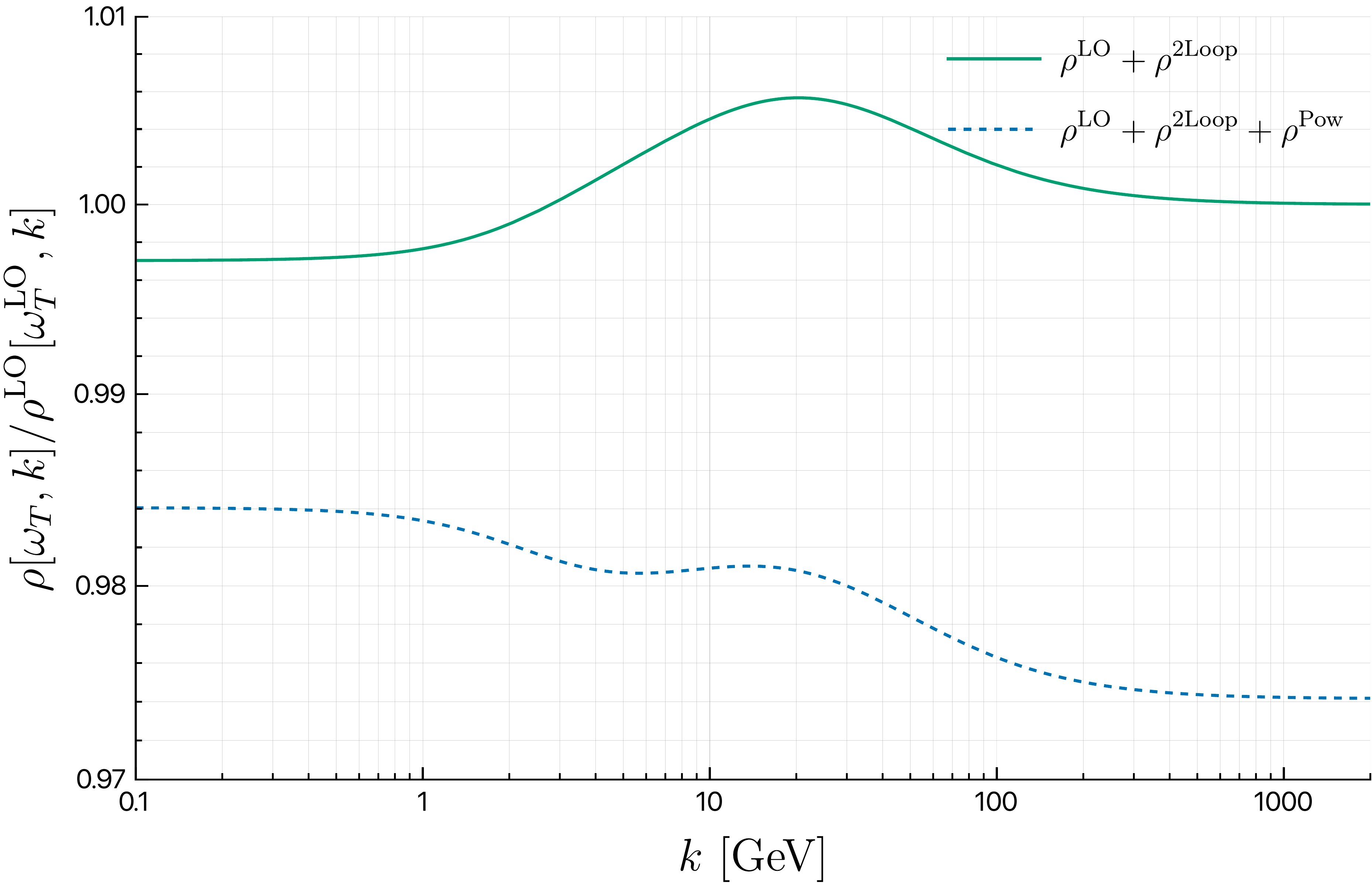}
	\end{center}
	\caption{The behaviour of the resummed residue at large $k$. The full line denotes the resummed one-loop residue evaluated in the resummed pole, and the dotted line also incorporates power corrections\te evaluated in the power-corrected pole. We use $\alpha=0.1$ and $\mu=T=10~\text{GeV}$ in both plots.}
	\label{fig:RhoTComp}
\end{figure}

Next, the residue. If we do not resum, the residue decreases faster than $k^{-1}$ for large $k$; which can hardly be interpreted as a physical excitation. If we resum our propagators, however, the residue does behave as $k^{-1}$. Though, note from figure \ref{fig:RhoTComp} that the residue lies slightly below $\rho^\text{res}=(2k)^{-1}$ once we incorporate power corrections. This is not unexpected, it simply means that our asymptotic $k\gg T$ fields are normalized differently from our $k^0= 0$ ones. In fact, from the power corrections we see that $A^T_{k^0 =0}=A^T_{k\gg T} \sqrt{Z}$, where $Z=(1+\frac{5e^2}{24\pi^2})^{-1}\approx 1-\frac{5e^2}{24\pi^2}\approx 0.973$.

What's more, it takes quite a while\te until $k\sim 10~T$\te before the pole and residue approach their asymptotic values. So the error of treating the pole as a massive excitation at $k\leq T$ is correspondingly large.

Finally, we can perform a similar analysis for the longitudinal propagator. Here we also find a pole for small $k^0,k \sim e T$, and at leading-order this pole preservers even for large~\cite{Pisarski:1989cs,LEBEDEV1990229} momenta: $k^0=k+\mathcal{O}(e^{-k^2/(e^2 T^2)})$. Yet contrary to the transverse pole, the longitudinal one should get Landau-damped into oblivion as $k$ approaches $T$~\cite{LEBEDEV1990229}. We confirm that this occurs once the self-energy has been resummed.
\vspace*{-0.5cm}
\subsection{Three-loop predictions}
While three-loop calculations lie a few years in the future, we can even now make an educated guess of their form. 

Indeed, a new Lorentz structure arose at two-loops because hard particles obtained a thermal mass. This gave terms proportional to $\Pi_2^{\mu \nu }$ where the proportionality constant depends on one-loop masses. We can use this as a guiding moonlight at higher orders. In particular, by expanding the one-loop self-energy we deduce a new Lorentz structure:
\begin{align}
\Pi_4^{\mu \nu}=&\int \frac{d\Omega_v}{4\pi}\left\{\left[\frac{(k^0)^3}{(v\cdot K)^3}-\frac{3}{2}\frac{(k^0)^2}{(v\cdot K)^2}+\frac{k^0}{v\cdot K}\right]v^\mu v^\nu \right.
\\&\left.+\left(v^\mu n^\nu+n^\mu v^\nu\right)\left[\frac{(k^0)^2}{(v\cdot K)^2}-\frac{3}{2}\frac{k^0}{v\cdot K}\right]+ n^\mu n^\nu\left[\frac{k^0}{v\cdot K}-\frac{1}{2}\right]\right\},
\end{align}
where we note that\footnote{This was guaranteed to happen since the resummed one-loop self-energy is conserved.} $\Pi_4^{\mu \nu}K_\nu=0$. If the mass comes from a fermion current we find that the factor multiplying $\Pi_4^{\mu \nu}$ is
\begin{align}
-\frac{21 \zeta(3)}{8\pi^4 T^2} m_F^4,
\end{align}
relative to the one-loop self-energy.

And if the mass comes from a boson current the factor is
\begin{align}
	\frac{3 \zeta(3)}{16\pi^2 T^2} m_{S/V}^4,
\end{align}

To be specific, for quantum electrodynamics (at $\mu_F=0$) we predict the three-loop term
\begin{align}
\Pi^{\mu \nu}_\text{three-loop}=\frac{7e^6 T^2 \zeta(3)}{128\pi^4} \Pi_4^{\mu \nu}+\ldots,
\end{align}
and for quantum chromodynamics we expect
\begin{align}
\Pi^{\mu \nu}_\text{three-loop}=\frac{\zeta(3) T^2 g_s^6}{16 \pi^4}\left\{\frac{7}{8} N_q T_F C_F^2-\frac{1}{36}C_A\left( C_A+N_q T_F\right)^2 \right\} \Pi_4^{\mu \nu}+\ldots
\end{align}
\vspace*{-1.5cm}

\section{Conclusions}
With the advent of gravitational-wave cosmology, the pressure is on the community to provide accurate predictions for dense objects like neutron stars. And because the strong coupling can be large ($\alpha_S\sim 1$), staying at leading order is not an option. To that end, this paper studied how gluons and photons are screened in hot and dense media. We extended the effective description of slowly varying fields\te hard thermal loops\te to two-loop precision. We used this to find next-to-leading-order corrections to the gluon's asymptotic mass and damping rate.

 For future work it would be quite interesting to also tackle soft quarks. This is important not only for dense objects, but also for particle production in hot plasmas where the asymptotic fermion mass plays a key role.

We end on the hope that even as two-loop investigations proceed, the community will go forth to the next order. And as we have argued, we can even now deduce the form of these three-loop corrections\te we leave it for the next generation to see whether this picture is clear or blurred.

\section*{Acknowledgements}
I am grateful to Tyler Gorda, Risto Paatelainen, Saga S\"appi, and Kaapo Sepp\"anen for engaging discussions; a critical read-through of the manuscript; and especially for performing cross-checks with their independent result. This work has been supported by the Swedish Research Council, project number VR:$2021$-$00363$.

\appendix

\section{Group invariants for simple gauge groups}\label{app:GroupInvariants}
The general result for hard thermal loops can be written in terms of various traces over generators. For simple gauge groups the result can be simplified by using the relations
\begin{align}
	&\text{Tr}\left[g_V^a g_V^b\right]=- C_A \delta^{ab},\quad \text{Tr}\left[g_F^a g_F^b\right]=2  \kappa T_F N_F \delta^{ab}, \quad \text{Tr}\left[g_S^a g_S^b\right]=-2 T_S N_S \delta^{ab},
	\\&g_V^{adc}g_V^{cef}g_V^{dfn}g_V^{enb}=\frac{1}{2}C_A^2 \delta^{ab}, \quad \text{Tr}\left[g_S^a g_S^b g_S^c g_S^c\right]=2 C_S T_S N_S \delta^{ab},  \nonumber
	\\& \text{Tr}\left[g_F^a g_F^b g_F^c g_F^c\right]=2 \kappa C_F T_F N_F \delta^{ab}, \quad  g^{ace}g^{bcd}\text{Tr}\left[g_F^d g_F^e\right]= 2 \kappa T_F T_A N_F \delta^{ab}, \nonumber
	\\& g _V^{aec}g_V^{bdc}\text{Tr}\left[g_S^d g_S^e\right] =-2 T_S T_A N_S \delta^{ab}, \quad \text{Tr}\left[g_F^a g_F^c g_F^b g_F^c\right]=\kappa T_F N_F \left(2 C_F- C_A\right) \delta^{ab}.
\end{align}
where $N_S$ and $N_F$ denotes the number of scalar and fermion representations respectively; $\kappa=1$ for Dirac and $\kappa=\frac{1}{2}$ for Weyl fermions; and for fundamental particles in $\mathrm{SU}(N)$
\vspace*{-0.5cm}\begin{align}
	C_{F/S}=\frac{N^2-1}{2 N}, \quad T_{F/S}=\frac{1}{2}, \quad C_A=T_A=N.
\end{align}
\vspace*{-1.5cm}

\section{Transverse and longitudinal self-energies}\label{app:TL}
The self energies are defined by
 \begin{align}
	\Pi_T=\frac{1}{d-1}\left[g_{\mu \nu}\Pi^{\mu \nu}+\frac{K^2}{k^2}\Pi^{00}\right], \quad \Pi_L=-\frac{K^2}{k^2} \Pi^{00}
\end{align}
To calculate the trace and $00$ component one can use
\begin{align}
	&\Pi_1^{00}=1- k^0 L[K], \quad g_{\mu \nu }\Pi_1^{\mu \nu}=-1,
	\\&\Pi_2^{00}=-1-\frac{(k^0)^2}{K^2}, \quad g_{\mu \nu }\Pi_2^{\mu \nu}=1+2 k^0 L[K],
	\\& \Pi_3^{00}= (1- k^0 L[K])^2, \quad g_{\mu \nu}\Pi_3^{\mu \nu}=-\frac{K^2}{k^2}(1- k^0 L[K])^2,
	\\& L[K]\equiv \frac{1}{2k}\log \frac{k^0+k+i \eta}{k^0-k+\eta}, \quad \eta=0^{+}.
\end{align}

\section{Results for quantum chromodynamics}\label{app:QCDRes}

\subsection{Self-energy}\label{app:QCDFieldShift}
The two-loop gluon self-energy is given by the sum of

\begin{align}
	&\!\begin{aligned}[t] 
			\Pi^{\mu \nu}_\text{two-loop}=&-\alpha_s^2 C_A^2 T^2 \left\{\frac{4}{3}\left[\log \frac{\mu e^\gamma }{4 \pi  T}+k^0 L[K] \right]\Pi_1^{\mu \nu}-\frac{2}{3} \Pi_2^{\mu \nu}\right\} \nonumber
		\\&-\alpha_s^2 N_q \left\{-\frac{2 \mu_F^2}{\pi^2}(2C_F T_F-C_A T_F)\Pi_3^{\mu \nu} +2 C_F T_F \left(\frac{\mu_F^2}{\pi ^2}+T^2\right)\Pi_2^{\mu \nu}\right\}
		\\&- \alpha_s^2 N_q \left\{\frac{4}{3} C_A T_F \left(\frac{3 \mu_F^2}{\pi ^2}+T^2\right) \left[\log \frac{\mu e^\gamma }{4 \pi  T}+k^0 L[K] \right] \Pi_1^{\mu \nu}-\frac{2}{3} C_A T_F \left(\frac{3 \mu_F^2}{\pi ^2}+T^2\right)\Pi_2^{\mu \nu}	\right\},
	\end{aligned}\\
&\text{and}\nonumber\\ 
	&\!\begin{aligned}[t] 
			\Pi^{\mu \nu}_\text{field-shift}=&-\frac{1}{9} \alpha_s^2 C_A^2T^2 \left(10 \log \frac{ \mu e^\gamma }{4 \pi  T}-1\right)\Pi^{\mu \nu}_1 \nonumber
		\\&+\alpha_s^2 \left\{\frac{4}{9}  N_q^2 T_F^2 \left(\frac{3 \mu_F^2}{\pi ^2}+T^2\right) \left(2 \log \frac{ \mu e^\gamma }{4 \pi  T}-1+4 \log (2)\right) \right\}\Pi^{\mu \nu}_1 \nonumber
		\\&-\frac{1}{9}\alpha_s^2 N_q C_A T_F  \left\{2 \left(\frac{15 \mu_F^2}{\pi ^2}+T^2\right)  \log \frac{ \mu e^\gamma }{4 \pi  T}+3 \left(T^2-\frac{\mu_F^2}{\pi ^2}\right)\right\}\Pi^{\mu \nu}_1 \nonumber
		\\&+ \frac{16}{9} \alpha_s^2 N_q C_A T_F T^2 \log (2) \Pi^{\mu \nu}_1.
	\end{aligned}
\end{align}
\subsection{Asymptotic gluon mass}\label{app:QCDMassRes}
The leading-order mass is
\begin{align}
	(m_g^2)_\text{LO}=\frac{1}{6}\left(C_A+N_q T_F\right) g_s^2 T^2+\frac{1}{2}N_q T_F \frac{\mu_F^2}{\pi^2} g_s^2,
\end{align}
while the next-to-leading order correction is given by the sum of
\begin{align}
&\!\begin{aligned}[t] 
m^2_\text{resum}=-\frac{ T_F N_q \alpha_s}{\pi}(m_q^2)_\text{LO} +\frac{C_A \alpha_s}{2 \pi}(m_g^2)_\text{LO} ,
\end{aligned}\\
&\!\begin{aligned}[t] 
		m^2_\text{ren}=&\frac{1}{9} \alpha_s^2 T^2 (11 C_A-4 N_q T_F)(C_A+N_q T_F)\log\frac{\mu e^\gamma}{4\pi T}
	\\&+\frac{\alpha_s^2 N_q T_F \mu_F^2 (11 C_A-4 N_q T_F)}{3 \pi ^2}\log\frac{\mu e^\gamma}{4\pi T},
\end{aligned}\\
&\!\begin{aligned}[t] 
m^2_\text{finite}=&-\alpha_s^2 \mu_F^2\frac{ N_q T_F C_A+4 N_q T_F (4 \log (2)-1)}{6 \pi ^2}
\\& -\frac{1}{18}\alpha_s^2 T^2 (C_A +N_q T_F) \left\{C_A+4 N_q T_F (4\log (2)-1)\right\},
\end{aligned}\\
\text{and}\\
&\!\begin{aligned}[t] 
	m^2_\text{power}=& -\alpha_s \frac{ (10 N_q T_F-11 C_A)}{12 \pi }(m_g^2)_\text{LO} 
	\\&+\frac{2}{3}\alpha_s^2\left\{T^2 C_A(C_A+N_q T_F)+\frac{3 \mu_F^2}{\pi^2}C_A N_q T_F\right\}.
\end{aligned}
\end{align}
\section{New Lorentz structure at finite chemical potential}\label{app:Pi3Lorentz}
As mentioned, terms proportional to $\Pi_3^{\mu \nu}$ only arises for a finite chemical potential. There are only two diagrams, both involving the fermion current.

\subsection{Vector contribution to the fermion current}\label{app:Pi3Terms}
The full vector contribution is given by (see \cite{Ekstedt:2023anj} for the details)
\begin{align}\label{eq:FermionCurrentGen}
	\Pi^\mu=g^{a N}_I g^{c J}_K g^{d L}_M\int_{PQ} F^{\mu}(P,Q)&\left\{ \delta^{M}_N \delta^K_L \Delta_{F,J}^{I}(P)\Delta_V^{cd}(Q) \left[ \Delta^R(P)\Delta^R(P+Q)+\Delta^A(P)\Delta^A(P+Q)\right]\right.\nonumber
	\\&\left. +\delta^{cd} \delta^{M}_N\Delta_{F,J}^{I}(P)\Delta^{K}_{F,L}(P+Q)\left[\Delta^R(P)\Delta^A(Q)+\Delta^A(P)\Delta^R(Q)\right] \right. \nonumber
	\\& \left.  +\delta^{M}_N \delta^{I}_J \Delta^{K}_{F,L}(P+Q)  \Delta^{cd}_V(Q)\left[\Delta^R(P)\Delta^A(P)\right]\right\},\nonumber
	\\& F^\mu=2(D-2)\left[(P+Q)^2 p^\mu-P^2 q^\mu -Q^2 p^\mu\right].
\end{align}
Here we are only interested in the term proportional to $\Delta_{F,J}^{I}(P)\Delta^{K}_{F,L}(P+Q)$. Now, in the background-field method propagators carry a soft momenta $K$. If the $\delta n_F\sim \vec{v}\cdot \vec{E}$ term comes from $\Delta_{F,J}^{I}(P)$ this is irrelevant since the soft momenta directly disappears into the resummed propagator. If however the field insertion comes from $\Delta^{K}_{F,L}(P+Q)$, the soft momenta has to travel through advanced and retarded propagators\footnote{These contributions are only possible with a finite chemical potential.}. So formally we have to use retarded and advanced propagators that depend on the background field. In practice however we are dealing with a single term, so it is much easier to just shift $\Delta^{R/A}(P)\rightarrow \Delta^{R/A}(P+K)$, and correspondingly $P\rightarrow P+K$ for the relevant term in the fermion trace. 

For the term in question we then get
\begin{align}
&g^{a N}_I g^{c J}_K g^{d L}_M \int_{PQ} F^\mu\delta^{M}_N\Delta_{F,J}^{I}(P)\Delta^{K}_{F,L}(P+Q)\left[\Delta^R(P)\Delta^A(Q)+\Delta^A(P)\Delta^R(Q)\right]
\\&=-\frac{ D-2}{16 \pi^4}g^{a N}_I g^{c J}_K g^{c L}_N (g^b \mu_F)^K_L \mu_{F,J}^I \Pi_3^{\mu\nu} A^b_\nu+\ldots,
\end{align}
Where $\ldots$ signify terms that are not proportional to $\Pi_3^{\mu \nu}$. To get this result we used
\begin{align}
&\int_p p^{-2} \left[n_F(p+\mu_F)-n_F(p-\mu_F)\right]=\frac{\mu_F}{4\pi^2},
\\& \int_p p^{-1}\left[n_F'(p+\mu_F)-n_F'(p-\mu_F)\right]=-\frac{\mu_F}{4\pi^2},
\end{align}
and (when $P^2=0$)
\begin{align}
\Delta^R(P+K)+\Delta^A(P+K)=-i\left[\frac{1}{P\cdot K}-\frac{K^2}{2 (P\cdot K)^2}+\ldots\right].
\end{align}

An analogous contribution arises from Yukawa corrections to the fermion current. One finds
\begin{align}
-\frac{1}{16\pi^4}\left[\mu_F g_F^a \mu_F g_F^b\right]^{I}_J (Y Y^c)^{J}_I\Pi_3^{\mu \nu}.
\end{align}

\section{Complete self-energy at next-to-leading order for any model}\label{app:FullResult}
We here give the general result for hard thermal loops. The first three subsection give the $\mu_F=0$ results, and the next one lists contributions that arise when $\mu_F\neq 0$. The full result is the sum of all following terms. See \cite{Ekstedt:2023anj} for the explicit Lagrangian.
\subsection{Two-loop contributions}
\begin{align}
&\frac{1}{24 \pi^2}T^2g_V^{adc}g_V^{cef}g_V^{dfn}g_V^{enb}\left\{\Pi_2^{\mu \nu}-2\left[\log\frac{\mu e^\gamma}{4\pi T}+k^0 L[K]\right]\Pi_1^{\mu \nu}\right\}\nonumber
\\&-\frac{1}{16 \pi ^2}T^2\text{Tr} g_F^c g_F^c g_F^a  g_F^b \Pi_2^{\mu \nu}+\frac{T^2}{48 \pi ^2}g_V^{ace}g_V^{bcd}\text{Tr}\left[g_F^d g_F^e\right]\Pi_2^{\mu \nu} -\frac{1}{8 \pi ^2}\text{Tr}\left[g_F^a M_F M_F^{\dagger} g_F^b\right]\Pi_2^{\mu \nu}\nonumber
\\&-\frac{1}{12\pi^2} g^{ace}_Vg^{bcd}_V\text{Tr}\left[g_F^d g_F^e\right]\left[\log\frac{\mu e^\gamma}{4\pi T}+k^0 L[K]\right]\Pi_1^{\mu \nu}-\frac{T^2}{192 \pi ^2} \text{Tr}\left[g_S^a g_S^b\right]_{jl}\lambda^{jlnn} \Pi_2^{\mu \nu}\nonumber
\\&-\frac{1}{8 \pi ^2}\text{Tr}\left[g_S^a g_S^b\right]_{ij}\mu^{ij} \Pi_2^{\mu \nu} + \frac{T^2}{32 \pi ^2}\text{Tr}\left[g_S^a g_S^b g_S^c g_S^c\right] \Pi_2^{\mu \nu}- \frac{T^2}{48\pi^2}g _V^{aec}g_V^{bdc}\text{Tr}\left[g_S^d g_S^e\right]\Pi_2^{\mu \nu}\nonumber
\\&-\frac{T^2}{32\pi^2}\left[g_F^a g_F^b\right]^{I}_J (Y Y^c)^{J}_I\Pi_2^{\mu \nu}
-\frac{T^2}{192\pi^2}\left[g_S^a g_S^b\right]_{ij}(Y Y^c+Y^c Y)^{ij}\Pi_2^{\mu \nu}
\\& + \frac{1}{24\pi^2} T^2g_V^{aec}g_V^{bdc}\text{Tr}\left[g_S^d g_S^e\right] \left[\log\frac{\mu e^\gamma}{4\pi T}+k^0 L[K]\right] \Pi_1^{\mu \nu}\nonumber
\end{align}

\subsection{Contributions from field-shifts}
These contributions arise if we make the $A^0 A^0$ propagator canonical when $k^0=0$; in a different conventions these terms can be omitted, but the full power-correction should then be used.
\begin{align}
&-\frac{1}{72 \pi^2}T^2 g_V^{adc}g_V^{cef}g_V^{dfn}g_V^{enb}\left[ 10\log\frac{\mu e^\gamma}{4\pi T}-1\right]\Pi_1^{\mu \nu}\nonumber
\\&+ \frac{ 1}{72 \pi ^2} T^2 \text{Tr}\left[g_F^a g_F^c\right]\text{Tr}\left[g_F^c g_F^b\right] \left[\log \frac{\mu e^\gamma}{4\pi T} -\frac{1}{2}+\log (4)\right]\Pi_1^{\mu \nu}\nonumber
\\&+ \frac{1}{288 \pi ^2} T^2 \left\{\text{Tr}\left[g_F^a g_F^c\right]\text{Tr}\left[g_V^c g_V^b\right] +\text{Tr}\left[g_V^a g_V^c\right]\text{Tr}\left[g_F^c g_F^b\right]  \right\} \left[ \log \frac{\mu e^\gamma}{4\pi T} +\frac{3}{2}-8 \log (2)\right]\Pi_1^{\mu \nu}\nonumber
\\&+\frac{1}{288 \pi ^2} T^2\text{Tr}\left[g_S^a g_S^c\right]\text{Tr}\left[g_S^c g_S^b\right]\left[ \log \frac{\mu e^\gamma}{4 \pi T}  +1\right]\Pi_1^{\mu \nu}
\\& -\frac{1}{72 \pi ^2} T^2\left\{\text{Tr}\left[g_S^a g_S^c\right]\text{Tr}\left[g_V^c g_V^b\right] +\text{Tr}\left[g_V^a g_V^c\right]\text{Tr}\left[g_S^c g_S^b\right]  \right\}\left[ \log \frac{\mu e^\gamma}{4\pi T}-3\right]\Pi_1^{\mu \nu}\nonumber
\\&-\frac{1}{576 \pi ^2} T^2\left\{\text{Tr}\left[g_S^a g_S^c\right]\text{Tr}\left[g_F^c g_F^b\right] +\text{Tr}\left[g_F^a g_F^c\right]\text{Tr}\left[g_S^c g_S^b\right]  \right\} \left[5 \log \frac{\mu e^\gamma }{4 \pi  T} -1+8 \log (2)\right] \Pi_1^{\mu \nu}\nonumber
\end{align}

\subsection{Contribution from power corrections}
For power-corrections it is easiest to not give the full Lorentz structures. For example, the $00$ component is
\begin{align}
\Pi_{00}^\text{Pow}=&-\text{Tr}\left[g_S^a g_S^b\right]\frac{k^2}{16 \pi^2}\left\{\frac{1}{3} \frac{(k^0)^2}{k^2} (k^0 L(K)-1)\right\}\nonumber
\\&-\text{Tr}\left[g_F^a g_F^b\right]\frac{k^2}{16 \pi^2}\left\{\frac{2}{3} k^0 \left(3-\frac{(k^0)^2}{k^2}\right)L(K)+\frac{2}{3}\frac{(k^0)^2}{k^2}\right\}
\\&-\text{Tr}\left[g_V^a g_V^b\right]\frac{k^2}{16 \pi^2}\left\{\frac{2}{3} k^0 \left(6-\frac{(k^0)^2}{k^2}\right)L(K)+\frac{2 (k^0)^2}{3 k^2}\right\},\nonumber
\end{align}
and the trace is
\begin{align*}
g^{\mu \nu }\Pi_{\mu \nu}^\text{Pow}=&\text{Tr}\left[g_S^a g_S^b\right]\frac{K^2}{16 \pi^2}\left\{k^0 L(K)-\frac{3}{3}\right\}+\text{Tr}\left[g_F^a g_F^b\right]\frac{K^2}{16 \pi^2}\left\{4 k^0 L(K)+\frac{4}{3}\right\}
\\&+\text{Tr}\left[g_V^a g_V^b\right]\frac{K^2}{16 \pi^2}\left\{10 k^0 L(K)+\frac{4}{3}\right\}.
\end{align*}

\subsection{Contributions with a chemical potential}\label{app:FiniteChemical}
\begin{align}\label{eq:FiniteMuF}
&-\frac{\log \frac{\mu e^\gamma}{4\pi T}}{24 \pi^2}g_V^{ace}g^{bcd}_V\text{Tr}\left[g^d_F\left(3\mu_F^2/\pi^2\right)g_F^e\right]\Pi_1^{\mu \nu}-\frac{1}{8\pi^4}\text{Tr}\left[g_F^a \mu_F g_F^c g_F^b \mu_F g_F^c\right]\Pi_3^{\mu \nu} \nonumber
	\\&+\frac{\log \frac{\mu e^\gamma}{4\pi T}-\frac{1}{2}+\log(4)}{72\pi^2}\text{Tr}\left[g_F^a\left(3\mu_F^2/\pi^2\right)g_F^c\right]\text{Tr}\left[g_F^c g_F^b\right]\Pi_1^{\mu \nu} \nonumber
	\\&-\frac{1}{16\pi^2}\text{Tr}\left[g^c_F g^c_F\left(3\mu_F^2/\pi^2\right)g_F^a g_F^b\right]\Pi_2^{\mu \nu}+\frac{1}{48\pi^2}g_V^{ace}g^{bcd}_V\text{Tr}\left[g^d_F\left(3\mu_F^2/\pi^2\right)g_F^e\right]\Pi_2^{\mu \nu}\nonumber
	\\&+\frac{15\log \frac{\mu e^\gamma}{4\pi T}-\frac{3}{2}}{288\pi^4}\left\{\text{Tr}\left[g^a_V g^c_V\right]\text{Tr}\left[g^c_F \mu_F^2 g^b_F\right]+\text{Tr}\left[g^a_F \mu_F^2 g^c_F\right]\text{Tr}\left[g^c_V g^b_V\right]\right\}
	\\&-\frac{1}{32\pi^4}\left[g_F^a \mu_F^2 g_F^b\right]^{I}_J (Y Y^c)^{J}_I\Pi_2^{\mu \nu}-\frac{1}{16\pi^4}\left[\mu_F g_F^a \mu_F g_F^b\right]^{I}_J (Y Y^c)^{J}_I\Pi_3^{\mu \nu} \nonumber
	\\&-\frac{1}{64 \pi^4}\left[g_S^a g_S^b\right]_{ij}(Y\mu_F^2 Y^c+Y^c  \mu_F^2Y)^{ij}\Pi_2^{\mu \nu}\nonumber
	\\& -\frac{ \log \frac{\mu e^\gamma }{4 \pi  T} +1}{192 \pi ^4}\left\{\text{Tr}\left[g_S^a g_S^c\right]\text{Tr}\left[g_F^c \mu_F^2 g_F^b\right] +\text{Tr}\left[g_F^a \mu_F^2 g_F^c\right]\text{Tr}\left[g_S^c g_S^b\right]  \right\} \Pi_1^{\mu \nu}.\nonumber
\end{align}
Finally, the extra power-correction is
\begin{align}
	&g_{\mu \nu }\delta \Pi^{\mu \nu}_{F}=\frac{4}{3}\text{Tr}\left[g_F^a  \overline{\aleph}_Fg_F^b\right]\frac{K^2}{16 \pi^2},\quad  \delta \Pi^{00}_F=-\frac{8}{3}\text{Tr}\left[g_F^a \overline{\aleph}_F g^b\right]\frac{k^2}{16 \pi^2},
\end{align}
where
\begin{align}
	\overline{\aleph}_F=-\left\{\frac{1}{2}\left[\psi\left(1/2-i \frac{\mu_F}{2\pi T}\right)+\psi\left(1/2+i \frac{\mu_F}{2\pi T}\right)\right]+\log(4)+\gamma\right\}.
\end{align}

\section{Expansion at the lightcone}\label{app:SelfEnergy}
When resumming our propagators we have to evaluate integrals of the type
\begin{align}\label{eq:NumIntegral}
F=\int dp p^{-1} \left(e^{p^0/T}\pm 1\right)^{-1} \frac{k^0 p^0}{2 p k} \log\frac{k^0 p^0+k p}{k^0 p^0-k p}, 
\end{align}
where $m_F^2$ denotes a thermal mass and  $p^0=\sqrt{p^2+m_F^2}$,. Away from the light-cone we are free to expand the integral in powers of $m_F^2$. However, close to the light-cone we have to keep the full $m_F^2$ dependence.  Since we are interested in power corrections, this integral multiplies $d-3=-2\epsilon$. This means that only $F\sim \frac{1}{\epsilon}$ terms contribute.

We will here be concerned with the $k^0,k\gg T\gg m_F,K$ limit. We see that the $p\sim  k$ region does not contribute as the Boltzmann factor kills all fluctuations, moreover, $F$ scales as $k^{-2}$ in this limit. For smaller momenta the integral behaves as (at least for the coefficient in-front of $\epsilon^{-1}$)
\begin{align}
F \approx \int dp p^{-1} \left(e^{p^0/T}\pm 1\right)^{-1}.
\end{align}
We can isolate the divergence to find 
\begin{align}
F=\pm \frac{1}{4\epsilon}+\ldots.
\end{align}
This is the same factor that we would find in front of $k^0 L[K]$ had we set $m_F=0$.
As such we can replace factors of $k^0 L[K]$ by $1$ when $k,k^0\gtrsim T$. 

\bibliographystyle{utphys}

{\linespread{0.6}\selectfont\bibliography{Bibliography}}

\end{document}